\documentstyle[12pt]{article}
\begin{document}
\newcommand{\bea}{\begin{eqnarray}}
\newcommand{\eea}{\end{eqnarray}}
\newcommand{\be}{\begin{equation}}
\newcommand{\ee}{\end{equation}}
\newcommand{\non}{\nonumber}
\newcommand{\ov}{\overline}
\global\parskip 6pt
\begin{titlepage}
\begin{center}
{\Large\bf State Sum Models and Simplicial}\\
\vskip .25in
{\Large\bf Cohomology}\\
\vskip .50in
Danny Birmingham \footnote{Supported by Stichting voor Fundamenteel
Onderzoek der Materie (FOM)\\
Email: Dannyb@phys.uva.nl}     \\
\vskip .10in
{\em Universiteit van Amsterdam, Instituut voor Theoretische Fysica,\\
Valckenierstraat 65, 1018 XE Amsterdam, \\
The Netherlands} \\
\vskip .50in
Mark Rakowski \footnote{Email: Rakowski@maths.tcd.ie}   \\
\vskip .10in
{\em School of Mathematics, Trinity College, Dublin 2, Ireland}  \\
{\em and}\\
{\em Dublin Institute for Advanced Studies, 10 Burlington Road, Dublin 4}\\
\end{center}
\vskip .10in
\begin{abstract}
We study a class of subdivision invariant lattice models based on
the gauge group $Z_{p}$, with particular emphasis on the four dimensional
example.
This model is based upon the assignment of field variables to both the $1$-
and $2$-dimensional simplices of the simplicial complex. The property of
subdivision invariance is achieved when the coupling parameter is quantized
and the field configurations are restricted to satisfy a type of mod-$p$
flatness condition. By explicit computation of the partition function
for the manifold $RP^{3} \times S^{1}$, we establish that the theory
has a quantum Hilbert space which differs from the classical one.
\end{abstract}
\vskip .25in
\begin{center}
ITFA-94-13\\
May 1994 - Revised 9/94
\end{center}
\end{titlepage}

\section{Introduction}
A series of $Z_{p}$ lattice models was introduced in \cite{BR}
which had the
very special property of being subdivision invariant. This means that the
partition function is insensitive to successively finer triangulations of
the underlying simplicial complex. One should regard this property as
the discrete analog of a continuum quantum field theory being metric
independent. The formulation of these models
involved the assignment of field variables to simplices of
various dimensions.
In three dimensions, only link based gauge fields are possible and that
theory reduced to the abelian Dijkgraaf-Witten model \cite{DW}.
A new four dimensional model was also
introduced which involved fields associated
to both $1$- and $2$-dimensional simplices of the simplicial complex.

A crucial element in securing the property of subdivision invariance
was to restrict the allowed field configurations to those satisfying
a certain ``flatness" condition; in addition, a quantization of
the coupling parameter was also necessary.
Solutions to the flatness conditions correspond to simplicial cohomology
classes of the underlying  complex $K$. The partition function
is a sum over these classes of a Boltzmann weight which captures a certain
kind of ``intersection'' of these field configurations. In our four
dimensional model, this intersection is between $H^{1}(K,Z_{p})$ and
$H^{2}(K,Z_{p})$.

Our aim here is to develop further the properties of this theory, and
specifically to establish ``non-triviality'' in four dimensions. By this
we mean that we have a topological field theory whose quantum Hilbert
space differs from the classical
one; it is simply a statement about the dependence of the theory on the
coupling parameter in the Boltzmann weight. This can be contrasted with
the Dijkgraaf-Witten model with gauge group $Z_{p}$, where there is
no distinction between the classical and quantum Hilbert spaces.
Recall that a topological
field theory in $d+1$ dimensions associates a Hilbert space
to each closed $d$-manifold. The $d+1$ dimensional theory then governs the
topology changing amplitudes between $d$-manifolds which appear on
the boundary. In \cite{DW}, such a model was constructed in three
dimensions and there the dimensions of the quantum Hilbert spaces for
various bounding Riemann surfaces were related to conformal field theory.
The novelty in our models is that one can study examples in four
and higher dimensions as well. These models should also prove
useful in the general classification programme of
topological field theory.

After reviewing some general properties, we consider in detail
the evaluation of the partition function on the manifold $RP^{3} \times
S^{1}$ which computes the dimension of the Hilbert space associated
to $RP^{3}$.  This can then be compared with the simple example
of $S^{3} \times S^{1}$. We relate our pedestrian formulation
of these theories with the Bockstein operator, and finally we present
some of the properties associated to $4$-manifolds with boundary,
including the behaviour under connected sum.

\section{General Formalism}

   A lattice model is based on a simplicial complex which combinatorially
encodes the topological structure of some manifold.
Let us recall some of the essential ingredients that are required
in such a formulation; we refer the reader to \cite{JM,Rot,Still}
for a more complete account.

   Let $V = \{ v_{i} \}$ denote a finite set of $N_{0}$ points which we will
refer to as the vertices of a simplicial complex. An ordered $k$-simplex
is an array of $k+1$ distinct vertices which we denote by,
\bea
[ v_{0},\cdots ,v_{k} ] \;\;.
\eea
It will usually be convenient to use simply the indices themselves
to label a given vertex when no confusion will arise, so the above
simplex is denoted more economically by $[0,\cdots ,k]$.
Pictorially, a $k$-simplex should be regarded as a point,  line segment,
triangle, or tetrahedron for $k$ equals zero through three respectively.
A simplex which is spanned by any subset of the vertices is called a face
of the original simplex.
An  orientation of a simplex is a choice of ordering of its vertices,
where we identify orderings that differ by an even permutation, but for
the models described here we will require an ordering of all vertices.
One then checks that the invariant we compute is actually independent
of the choice made in vertex ordering.

The boundary operator $\partial$ on the ordered simplex
$\sigma=[v_{0},\cdots,v_{k}]$  is defined by,
\bea
\partial \,\sigma = \sum^{k}_{i=0}\; (-1)^{i}\, [v_{0},\cdots,
\hat{v}_{i},\cdots,v_{k}]\;\;,
\eea
where the `hat' indicates a vertex which has been omitted. It
is easy to show that the composition of boundary operators is zero;
$\partial^{2} = 0$.

We model a closed $n$-dimensional manifold as a collection
$K= \{ \sigma_{i} \}$ of $n$-simplices
constructed from the set of vertices $V$, subject to a few technical
conditions. Most importantly, every $(n-1)$-face of any given $n$-simplex
appears as an $(n-1)$-face of precisely two different $n$-simplices in
the collection $K$. One thinks of the $n$-simplices then as glued
together along $(n-1)$-faces.
In order to ensure that the simplicial
complex represents a manifold, we require
the ``link" of each vertex to be a combinatorial $(n-1)$-dimensional
sphere. We refer the reader to \cite{JM,Still} for a more
complete discussion of this condition.

The dynamical variables in the theories we construct will be
objects which assign an element in the cyclic group $Z_{p} = Z / pZ$,
which we represent as the set of integers,
\bea
\{ 0,\cdots , p-1 \} \;\; ,
\eea
to ordered simplices of some specified dimension. We call these dynamical
variables $k$-colours with coefficients in $Z_{p}$, and denote
the evaluation of some $k$-colour $B^{(k)}$ on the ordered $k$-simplex
$[0,\cdots ,k]$ by
\bea
< B^{(k)}, [0,\cdots ,k] > = B_{0 \cdots k} \in Z_{p} \;\; .
\eea
The superscript $(k)$ will usually be omitted when its value is clear
from context.
It is important to note that we are assigning a $Z_{p}$
element in a way which depends on the ordering of vertices in the simplex;
we do not have the rule $B^{(1)}_{01} = - B^{(1)}_{10}$, for example.
Instead, we shall assume that,
\bea
B^{(1)}_{10} = - B^{(1)}_{01} \;\;mod\;\; p \;\; ,
\eea
and similarly extend this to a $k$-colour for odd permutations of the
vertices. The case closest to conventional lattice gauge theory is
where a 1-colour variable is assigned to every 1-simplex in the complex.

The coboundary operator $\delta$ acts on the dynamical variables as follows.
Given a $(k-1)$-colour, an application of the coboundary operator
produces an integer in $Z$, when evaluated on an ordered $k$-simplex, namely
\bea
< \delta B^{(k-1)} , [0,\cdots , k] > &=& < B, \partial [0,\cdots , k]>
\nonumber \\
&=& B_{123\cdots k} - B_{023 \cdots k} + B_{013\cdots k} - \cdots \;\; .
\eea
We must emphasize that the above sum of integers is not taken with
modular $p$ arithmetic; it is simply an element in $Z$.
In cases where we will need to take some combination mod-$p$,
we will put those terms between square brackets, so for example,
\bea
[ a + b ] = a + b \;\; mod\,\, p \;\; .
\eea

There is also a cup product operation on colours which takes a $k$-colour
$B^{(k)}$ and a $l$-colour $C^{(l)}$ and gives an integer in $Z$
when evaluated on a $(k+l)$-ordered simplex:
\bea
< B\cup C, [0,\cdots ,k+l ] > = B_{0\cdots k} \cdot C_{k\cdots k+l} \;\; .
\eea
Note once again that this product is in $Z$ and the value is not taken
mod-$p$.

Let us now put these ingredients together and define our theories. First,
we must be given some oriented simplicial complex $K$ which we take to
represent a manifold of dimension $n$. One then has some collection
of $n$-simplices defined up to orientation. Take the vertex set of
this complex and give it an ordering. This is done arbitrarily and we
will have to show that our construction is independent of this choice,
see for example \cite{DW,Y1,CY}.
Now we can write down an ordered collection of the $n$-simplices; each
of the simplices is written in ascending order and a sign in front of that
simplex indicates whether that ordering is positively or negatively
oriented with respect to the orientation of the complex K. Let us denote
this ordered set of $n$-simplices by $K^{n}$,
\bea
K^{n} = \sum_{i} \;\; \epsilon_{i}\, \sigma_{i}\;\; ,
\eea
where the index $i$ runs over the ordered $n$-simplices $\sigma_{i}$ and
$\epsilon_{i}$ is a sign which indicates the orientation. We will
assign a Boltzmann weight $W[K^{n}]$ to $K^{n}$ by taking a
product of factors,
one for every $n$-simplex,
\bea
W[K^{n}] = \prod_{i} \;\; W[\sigma_{i}]^{\epsilon_{i}}\;\; .
\eea
Each of the individual factors is a nonzero complex number and will be some
function of the colours. The details of which colours we use and how the
function is defined will depend on the particular model. Finally, the
partition function, which we will require to be a
combinatorial invariant,
is defined to be a quantity which is proportional to the sum
of the Boltzmann weights over all colourings,
\bea
Z = \frac{1}{|G|^{f(N)}}\sum_{colours} \;\; W[K^{n}] \;\; .\label{z}
\eea
Here $|G|$ is the order of the gauge group and $f(N)$ is a function
of the number of simplices of various dimensions. This function will
be fixed for any given theory by scaling considerations. In the four
dimensional model to be discussed next, $f(N) = N_{1}$ where $N_{1}$
is the number of 1-simplices in the simplicial complex. In the three
dimensional Dijkgraaf-Witten model (based on a single 1-colour
field), as formulated in \cite{BR}, it
is equal to the number of vertices $N_{0}$.

\section{State Sum Model in Four Dimensions}

Let us now turn our attention to the four dimensional model of interest.
This model is based upon the assignment of field variables to both the
$1$- and $2$-dimensional simplices of the simplicial complex.
The Boltzmann weight of an ordered $4$-simplex $[0,1,2,3,4]$ is defined by:
\bea
W[[0,1,2,3,4]] &=& \exp\{\beta <B^{(2)} \,\cup \,
\delta A^{(1)}, [0,1,2,3,4]>\}  \non\\
&=&  \exp\{\beta B_{012}\, ( A_{23} + A_{34} - A_{24})\}
\;\;,     \label{bw}
\eea
where $B^{(2)}$ and $A^{(1)}$ are $2$- and $1$-colour fields, respectively.
Here, $\beta$ is a complex number which is as yet unrestricted;
we shall also find it convenient to use the scale factor $s = \exp[\beta]$.
The first
item on the agenda is to demonstrate that the Boltzmann weight
defines a theory which is subdivision invariant. As we shall see, this
requirement will enforce a quantization of the coupling parameter, and
lead to a restriction on the allowed colour configurations.
In order to establish the property of subdivision invariance, it is
sufficient to show that the Boltzmann weight is invariant under
a set of moves known as the Alexander moves \cite{Alex}. Equivalently,
for the case of closed manifolds, we can establish invariance by
examining the behaviour under a set of $(k,l)$ moves \cite{Pachner},
which we now recall.

{\em The $(k,l)$ Moves:}\\

In the four dimensional case of interest here, we have
five $(k,l)$ moves, with $k=1,\cdots,5$, and $k+l=6$.
It suffices to consider the first three cases; the $(4,2)$ and $(5,1)$
moves are inverse to the $(2,4)$ and $(1,5)$ moves, respectively.

The $(1,5)$ move: \\

This is described by adding a new vertex $x$ to the centre of the
$4$-simplex $[0,1,2,3,4]$, and linking it to the other $5$ vertices.
The original $4$-simplex is then replaced by an assembly of five
$4$-simplices, written symbolically as:
\begin{eqnarray}
[0,1,2,3,4] &\rightarrow&
[x,1,2,3,4] - [x,0,2,3,4] + [x,0,1,3,4] \non\\
&-& [x,0,1,2,4] + [x,0,1,2,3]\;\;.
\end{eqnarray}
This move is also known as an Alexander move of type $4$. Note also
that we declare the new vertex $x$ to be the first in the total ordering
of all vertices.

The $(2,4)$ move: \\

In this case, two $4$-simplices which share a common $3$-simplex
$[0,1,2,3]$ are replaced by four $4$-simplices sharing a common
$1$-simplex $[x,y]$:
\bea
& & [x,0,1,2,3] - [y,0,1,2,3] \rightarrow \non \\
& & [x,y,1,2,3] - [x,y,0,2,3] + [x,y,0,1,3] -[x,y,0,1,2]\;\;.
\eea
Again, we place the new vertices $x,y$ at the beginning of the vertex list.

The $(3,3)$ move:
\bea
& & [y,z,0,1,2] - [x,z,0,1,2] + [x,y,0,1,2] \rightarrow \non \\
& & [x,y,z,1,2] - [x,y,z,0,2] + [x,y,z,0,1]\;\;.
\eea
We note that the $2$-simplex $[0,1,2]$ is common to the left hand side, with
$[x,y,z]$ being common to the right.

For the case of the $(1,5)$ move, one finds that the
Boltzmann weights before and after subdivision are related by:
\bea
& &W[[0,1,2,3,4]] s^{-<\delta B \,\cup\, \delta A, \, [x,0,1,2,3,4]>}
= W[[x,1,2,3,4]] \label{6W}   \\
& &W[[x,0,2,3,4]]^{-1}\, W[[x,0,1,3,4]]\,
W[[x,0,1,2,4]]^{-1}\, W[[x,0,1,2,3]]\;\;.\non
\eea
It is immediately evident that the Boltzmann weight is not
generally invariant
under this move, due to the presence of the added ``insertion"
on the left hand side of (\ref{6W}). Our task is therefore to trivialize
this unwanted insertion factor, and this can indeed be achieved by imposing
a restriction on the sum over colourings and on the parameter $\beta$.
Subdivision invariance of this
four dimensional theory is now guaranteed by imposing quantization of
the coupling $s^{p^{2}} = 1$, as well as
a restriction of the colourings to those satisfying the conditions
\bea
[ \delta B^{(2)} ] = [ \delta A^{(1)}] = 0\;\; .\label{ba4d}
\eea
We shall refer to these restrictions as ``flatness" conditions.
For example, on the $2$-simplex $[0,1,2]$, we have the restriction
on the $1$-colour field
\be
[\delta A]_{012} \equiv [A_{12} - A_{02} + A_{01}] = 0\;\;.
\ee
As a reminder, we note that this particular equation can also be
written as
\be
[A_{01} + A_{12}] = A_{02}\;\;.
\ee
On the $3$-simplex $[0,1,2,3]$, the
restriction on the $2$-colour takes the form:
\be
[\delta B]_{0123} \equiv [B_{123} - B_{023} + B_{013} - B_{012}] = 0 \;\;.
\ee
With these restrictions, the product $\delta B \cup \delta A$ is clearly
a multiple of $p^{2}$ and the above insertion becomes unity. The resulting
identity involving the six Boltzmann weight factors shall be referred to
as the $6W$ identity. It is worth pointing out that invariance is
achieved here without the necessity of summing over the additional
configurations attached to the vertex $x$.

It requires little extra work to complete the
demonstration of subdivision invariance. One first notes that the remaining
$(k,l)$ moves also involve six Boltzmann weight factors, and it is easy to
see that the $6W$ identity is also a statement of invariance under
the $(2,4)$ and $(3,3)$ moves.

The subdivision invariant Boltzmann weight for the $4$-simplex
$[0,1,2,3,4]$
is given by:
\be
W[[0,1,2,3,4]] = \exp\{\frac{2 \pi i k}{p^{2}}\,
B_{012}\,( A_{23}\,+\, A_{34} \,-\,
[ A_{23} + A_{34}] )\}\;\;, \label{sdibw}
\ee
with $k \in \{0,1,\cdots,p-1\}$.

At this point, we can reveal that each of the colour fields enjoys a
local gauge invariance.
The gauge transformation of the A field defined on the ordered
$1$-simplex $[0,1]$ is defined by:
\be
A^{\prime}_{01} = [A - \delta \omega]_{01}
= [A_{01} - \omega_{1} + \omega_{0}]\;\;,
\ee
where $\omega$ is a $0$-colour field defined on the vertices of the complex.
For the $2$-colour field $B$ defined on the
ordered $2$-simplex $[0,1,2]$, we have  a gauge transformation given by:
\be
B'_{012} = [B - \delta \lambda]_{012} =
[B_{012} - \lambda_{12} + \lambda_{02} - \lambda_{01}]\;\;,
\ee
where $\lambda$ is a $1$-colour defined on $1$-simplices.
Our task now is to show that the Boltzmann weight for the case of
a closed simplicial complex is invariant with respect to independent
gauge transformations of the $A$ and $B$ fields.
As we shall see, invariance of the theory under the above transformations
is not manifest, but requires both the quantization of the coupling
parameter, together with the restriction on the allowed field configurations.

Under the transformation of B, one finds that
\be
s^{B^{\prime} \cup \delta A} =
s^{B \cup \delta A} s^{- \delta \lambda \cup \delta A} =
s^{B \cup \delta A} s^{-\delta (\lambda \cup \delta A)}\;\;,
\label{bwgt}
\ee
where the first equality uses the fact that $\delta A$ is an integer
multiple of $p$ due to the flatness constraint, and that $s$ is a
$p^{2}$-root of unity.
Hence, the Boltzmann weight is invariant up to a total boundary term
and the product of all these cancels for a closed oriented complex.
To demonstrate invariance under the $A$ field transformation,
one first notes the simple identity
\be
s^{B \cup \delta A} = s^{-\delta B \cup A} s^{\delta(B \cup A)} \;\;.
\ee
Invariance then follows immediately by the above argument.

As discussed in the previous section, the Boltzmann weight is initially
defined for a specific ordering of the vertex set.
We recall here a simple argument presented in \cite{Felder} which
can be used to
verify that the value of the partition function is independent
of this choice.

Let $V = \{v_{0},...,v_{N_{0}-1}\}$ be the vertex set of the complex,
$I$ the index set $I = \{0,...,N_{0}-1\}$,
and define a vertex ordering to be a map $f : V \rightarrow I$.
Clearly, if $f^{\prime}$ is a different vertex ordering, then
the composition $f^{\prime} \circ f^{-1}$ is a permutation
on the set $I$. Furthermore, to each permutation there is a
corresponding vertex ordering. Since  any permutation of $I$ can be
decomposed as a product of transpositions of consecutive numbers,
it suffices to show that the Boltzmann weight is invariant when
two consecutive values of the ordering $f$ are
permuted. Our task is therefore to show that the Boltzmann weights defined
with an ordering $f$, and $f^{\prime} = \pi \circ f$, coincide. Here,
the permutation $\pi$ is defined by $\pi(j) = j + 1, \pi(j+1) = j$
for some $j$, and $\pi(i) = i$ if $i \neq \{j,j+1\}$.

If $j$ and $j+1$ label vertices which do not bound a $1$-simplex,
then the Boltzmann weight is clearly invariant.
This follows because $j$ and $j+1$ are simply dummy variables
which can be freely exchanged, without affecting the orientation
of any individual $4$-simplex in the complex.

In order to establish invariance when the vertices labelled $j$ and $j+1$
bound a $1$-simplex, we recall the definition of an
Alexander move of type $1$. Given an ordered  $4$-simplex
$[v_{0},v_{1},v_{2},v_{3},v_{4}]$, we introduce an additional vertex
$x$ at the centre of the $1$-simplex $[v_{0},v_{1}]$, giving rise to the move
\be
[v_{0},v_{1},v_{2},v_{3},v_{4}] \rightarrow
[x,v_{1},v_{2},v_{3},v_{4}] - [x,v_{0},v_{2},v_{3},v_{4}]\;\;.
\ee
Since we have shown that the Boltzmann weight is invariant under the
$(k,l)$ moves, it is equivalently invariant under all Alexander moves.
Thus we are free to perform an Alexander move of type $1$
on the $1$-simplex with vertices labelled by $j$ and $j+1$.
This has the effect
that these vertices no longer bound a $1$-simplex, and by the above argument
$j$ and $j+1$ can then be interchanged leaving the Boltzmann weight
invariant.
In order to recover the original complex with the permuted vertex ordering,
one simply performs the inverse Alexander move of type $1$.

We have already shown that the Boltzmann weight is invariant under all
the $(k,l)$ subdivision moves. However, recall that to achieve subdivision
invariance, we are required to restrict the allowed field configurations
to those satisfying the appropriate flatness conditions. This is effected
in the state sum through the insertion of a set of delta functions which
implement the required restrictions.
It remains to check the behaviour of these
delta functions under the $(k,l)$ moves. As we shall see, the true
subdivision invariant partition function is given by including a certain
scaling factor, as discussed in relation to equation (\ref{z}). This
takes into account the redundancy in the assembly of delta
functions which are present under subdivision.

In order to determine the correct scaling factor, we need to examine the
behaviour of both the $A$ and $B$ delta functions with respect to
the $(k,l)$ moves. If we denote by $\Delta N_{i}$ the increase in the
number of $i$-simplices due to a $(k,l)$ move, then it is
straightforward to check that under the $(1,5)$ move we have:
\bea
\Delta N_{0} &=& 1\non\\
\Delta N_{1} &=& 5\non\\
\Delta N_{2} &=& 10\non\\
\Delta N_{3} &=& 10\non\\
\Delta N_{4} &=& 4\;\;.
\eea
The changes under the $(2,4)$ move are given by
\bea
\Delta N_{0} &=& 0\non\\
\Delta N_{1} &=& 1\non\\
\Delta N_{2} &=& 4\non\\
\Delta N_{3} &=& 5\non\\
\Delta N_{4} &=& 2\;\;,
\eea
and of course under the $(3,3)$ move we have $\Delta N_{i} = 0$, for all
$i$.

Let us now consider the behaviour of the $B$ delta functions under
subdivision. We will first collect some formulas and then put the
results together to determine the form of the scaling factor $f(N)$
referred to in equation (\ref{z}).
If we denote the additional ten $B$ fields present
after a $(1,5)$ move by:
\be
I = \{B_{x01},B_{x02},B_{x03},B_{x04},B_{x12},
B_{x13},B_{x14},B_{x23},B_{x24},B_{x34}\}   \;\;,
\ee
then one readily finds that summation over these fields yields the
result
\bea
\frac{1}{|G|^{4}}\sum_{I}
 \delta([\delta B]_{x012}) \delta([\delta B]_{x013})
    \delta([\delta B]_{x014}) \delta([\delta B]_{x023})
    \delta([\delta B]_{x024})& &  \non\\
 \delta([\delta B]_{x034}) \delta([\delta B]_{x123})
    \delta([\delta B]_{x124}) \delta([\delta B]_{x134})
    \delta([\delta B]_{x234})& &  \non\\
    \delta([\delta B]_{0123}) \delta([\delta B]_{0124})
    \delta([\delta B]_{0134}) \delta([\delta B]_{0234})
    \delta([\delta B]_{1234})& &   \non\\
\non\\
= \delta([\delta B]_{0123}) \delta([\delta B]_{0124})
    \delta([\delta B]_{0134}) \delta([\delta B]_{0234})
    \delta([\delta B]_{1234})& &\;\;.
\eea
Here, the assembly of delta functions on the right and left hand sides
above represent the situation before and after subdivision.
We specify that the modulo-$p$ delta function is defined by
\bea
\delta (x) =   \left\{  \begin{array}{ll}
      1  & \mbox{if $x$ = 0 mod-$p$} \\
      0                               & \mbox{otherwise.}  \\
      \end{array}
      \right.
\eea

For the case of the $(2,4)$ move, one finds that summation over the
additional four $B$ fields
\be
I = \{B_{xy0},B_{xy1},B_{xy2},B_{xy3}\}\;\;,
\ee
produces the result:
\bea
\frac{1}{|G|} \sum_{I}
    \delta([\delta B]_{xy01}) \delta([\delta B]_{xy02})
    \delta([\delta B]_{xy03}) \delta([\delta B]_{xy12})
    \delta([\delta B]_{xy13})& &   \non\\
    \delta([\delta B]_{xy23}) \delta([\delta B]_{x012})
    \delta([\delta B]_{x013}) \delta([\delta B]_{x023})
    \delta([\delta B]_{x123})& &  \non\\
    \delta([\delta B]_{y012}) \delta([\delta B]_{y013})
    \delta([\delta B]_{y023}) \delta([\delta B]_{y123})
    \phantom{\delta([\delta B]_{0123})} & &\non\\
\non \\
=   \delta([\delta B]_{x012})
    \delta([\delta B]_{x013}) \delta([\delta B]_{x023})
    \delta([\delta B]_{x123}) \delta([\delta B]_{y012}) & &   \non\\
    \delta([\delta B]_{y013}) \delta([\delta B]_{y023})
    \delta([\delta B]_{y123}) \delta([\delta B]_{0123})
    \phantom{\delta([\delta B_{0123}])}& &\;\;.
\eea

Turning now to the delta function insertions for the $A$ field, we
proceed in a similar manner.
The additional $A$ fields present after a $(1,5)$ move are:
\be
I = \{A_{x0},A_{x1},A_{x2},A_{x3},A_{x4}\}  \;\;.
\ee
One verifies that the following relation holds:
\bea
\frac{1}{|G|}\sum_{I}
    \delta([\delta A]_{x01}) \delta([\delta A]_{x02})
    \delta([\delta A]_{x03}) \delta([\delta A]_{x04})
    \delta([\delta A]_{x12}) & & \non\\
    \delta([\delta A]_{x13}) \delta([\delta A]_{x14})
    \delta([\delta A]_{x23}) \delta([\delta A]_{x24})
    \delta([\delta A]_{x34}) & &   \non\\
    \delta([\delta A]_{012}) \delta([\delta A]_{013})
    \delta([\delta A]_{014}) \delta([\delta A]_{023})
    \delta([\delta A]_{024}) & &  \non\\
    \delta([\delta A]_{034}) \delta([\delta A]_{123})
    \delta([\delta A]_{124}) \delta([\delta A]_{134})
    \delta([\delta A]_{234}) & &  \non\\
\non \\
=   \delta([\delta A]_{012}) \delta([\delta A]_{013})
    \delta([\delta A]_{014}) \delta([\delta A]_{023})
    \delta([\delta A]_{024}) & &  \non\\
    \delta([\delta A]_{034}) \delta([\delta A]_{123})
    \delta([\delta A]_{124}) \delta([\delta A]_{134})
    \delta([\delta A]_{234}) & & \;\;.
\eea

Finally, we treat the $(2,4)$ move for the $A$ field. There is a single
additional $A$ field $I = \{A_{xy}\}$ which is present after subdivision.
Summation  over this field produces the result:
\bea
\sum_{I}
    \delta([\delta A]_{xy0}) \delta([\delta A]_{xy1})
    \delta([\delta A]_{xy2}) \delta([\delta A]_{xy3})
    \delta([\delta A]_{012}) & & \non\\
    \delta([\delta A]_{013}) \delta([\delta A]_{023})
    \delta([\delta A]_{123}) \delta([\delta A]_{x01})
    \delta([\delta A]_{x02}) & & \non\\
    \delta([\delta A]_{x03}) \delta([\delta A]_{x12})
    \delta([\delta A]_{x13}) \delta([\delta A]_{x23})
    \delta([\delta A]_{y01}) & & \non\\
    \delta([\delta A]_{y02}) \delta([\delta A]_{y03})
    \delta([\delta A]_{y12}) \delta([\delta A]_{y13})
    \delta([\delta A]_{y23}) & & \non\\
\non \\
=   \delta([\delta A]_{012}) \delta([\delta A]_{013})
    \delta([\delta A]_{023}) \delta([\delta A]_{123})
    \delta([\delta A]_{x01}) & &  \non\\
    \delta([\delta A]_{x02}) \delta([\delta A]_{x03})
    \delta([\delta A]_{x12}) \delta([\delta A]_{x13})
    \delta([\delta A]_{x23})  & &     \non\\
    \delta([\delta A]_{y01}) \delta([\delta A]_{y02})
    \delta([\delta A]_{y03}) \delta([\delta A]_{y12})
    \delta([\delta A]_{y13}) & & \non\\
    \delta([\delta A]_{y23}) \phantom{\delta([\delta A]_{y01})
    \delta([\delta A]_{y02}) \delta([\delta A]_{y01})
    \delta([\delta A]_{y02})} & & \;\;.
\eea

We can now establish the correctly scaled subdivision invariant
partition function by combining
the previous results. Under the $(1,5)$ move, we see that a factor $|G|^{5}$
must be accounted for in the combined $A$ and $B$ sectors, and a factor
of $|G|^{1}$ under the $(2,4)$ move. But this is precisely how the number
of 1-simplices changes under these moves. If the partition
function of (\ref{z})
is chosen to have $f(N)= N_{1}$, then  it defines a
subdivision invariant quantity. Specifically, we have
\be
Z = \frac{1}{|G|^{N_{1}}}\; \sum_{flat} \; W[K^{n}]  \;\;, \label{z4dim}
\ee
where we denote the set of allowed colours satisfying the flatness
conditions by $flat$.
Clearly, at the trivial $s = 1$ root of unity
($k = 0$ in equation (\ref{sdibw})), the value of the partition
function simply counts the number of solutions to
the flatness conditions. Our main goal is in achieving interesting
behaviour at the non-trivial roots of unity where different phase factors
can occur.

\section{Evaluation of the Partition Function}

   The models described in the preceding sections require that a space
be presented as a simplicial complex for their formulation. It is
clear that one can only hope for a non-trivial partition function -
one in which the phases (Boltzmann weights) are not all unity - when
the field configurations we sum over are sufficiently interesting. This
means that we need solutions to the flatness equations (\ref{ba4d})
which would not be solutions
in the ``strong sense'' if the mod-$p$ brackets had been removed.
Perhaps the simplest example in four dimensions is the space
$RP^{3} \times S^{1}$, and we will give here a rather detailed
exposition of its simplicial description.

   Let us begin by presenting an economical simplicial complex for
the manifold $RP^{3}$. A complex with a minimal number of 11 vertices
has been given in \cite{Brehm}, and we label its vertices by elements
in the set $\{0,1,...,9,a\}$.  The complex is fully determined by
specifying the 3-simplices; these are 40 in number and are given
explicitly by,
\bea
& &+ [0,2,9,a] + [0,2,3,9] - [0,2,3,7] - [0,2,7,a] + [0,5,7,a] \\
& &- [0,4,5,7] + [0,1,4,5] + [0,1,3,4] - [0,1,3,9] + [0,1,6,9] \nonumber\\
& &+ [0,1,5,6] - [0,5,6,a] - [0,6,9,a] + [4,6,9,a] + [4,6,7,9] \nonumber\\
& &- [4,5,7,9] + [5,7,8,9] - [5,7,8,a] + [1,7,8,a] - [1,7,8,9] \nonumber\\
& &- [1,6,7,9] + [1,2,6,7] + [1,2,5,6] + [1,2,4,5] - [1,2,4,a] \nonumber\\
& &+ [1,3,4,a] - [1,3,8,a] + [1,3,8,9] + [3,5,8,9] - [2,3,5,9] \nonumber\\
& &+ [2,4,5,9] + [2,4,9,a] - [3,5,8,a] + [3,5,6,a] - [3,4,6,a] \nonumber\\
& &+ [3,4,6,7] + [2,3,6,7] + [2,3,5,6] + [1,2,7,a] - [0,3,4,7] \;\;
,\nonumber
\eea
where the signs denote the relative orientations of each simplex. Of course,
the lower dimensional simplices are given by all those which appear as
subsimplices in the above list. This complex contains 51 1-simplices
and 80 2-simplices in addition to the 11 vertices and 40 3-simplices
already tabulated. The Euler number is zero as required for a closed
3-manifold.
One also easily checks that the boundary of
the above complex vanishes and that these 3-simplices are glued
together along paired 2-simplices.

   Constructing the complex for $RP^{3}\times S^{1}$ is straightforward.
We begin by imagining the above complex of 40 3-simplices displayed
horizontally.
To each of those we add a new vertex and construct a vertical tower beneath
which contains a total stack of 12 4-simplices; this is the $S^{1}$
direction which gets glued to the top along a common 3-simplex. So, for
example, the tower beneath $+ [0,1,3,4]$ is given explicitly by,
\bea
& & + [0\phantom{''} , 1\phantom{''} , 3\phantom{''} , 4\phantom{''} , 0'
\phantom{'}]  \label{tower}\\
& & - [1\phantom{''} , 3\phantom{''} , 4\phantom{''} , 0'\phantom{'},
1'\phantom{'}] \nonumber \\
& & + [3\phantom{''} , 4\phantom{''} , 0'\phantom{'}, 1'\phantom{'},
3'\phantom{'}] \nonumber \\
& & - [4\phantom{''} , 0'\phantom{'}, 1'\phantom{'}, 3'\phantom{'},
4'\phantom{'}] \nonumber \\
& & + [0'\phantom{'}, 1'\phantom{'}, 3'\phantom{'}, 4'\phantom{'},
0''] \nonumber \\
& & - [1'\phantom{'}, 3'\phantom{'}, 4'\phantom{'}, 0'', 1''] \nonumber \\
& & + [3'\phantom{'}, 4'\phantom{'}, 0'', 1'', 3''] \nonumber \\
& & - [4'\phantom{'}, 0'', 1'', 3'', 4''] \nonumber \\
& & + [0'', 1'', 3'', 4'', 0\phantom{''} ] \nonumber \\
& & - [1'', 3'', 4'', 0\phantom{''} , 1\phantom{''} ] \nonumber \\
& & + [3'', 4'', 0\phantom{''} , 1\phantom{''} , 3\phantom{''} ] \nonumber \\
& & - [4'', 0\phantom{''} , 1\phantom{''} , 3\phantom{''} ,
4\phantom{''} ] \;\; . \nonumber
\eea
One sees that for each vertex $x$ in the original complex for $RP^{3}$,
two new vertices $x'$ and $x''$ are required in this presentation of
$RP^{3}\times S^{1}$. The total vertex set now is then,
\bea
\{ 0, ... , a, 0', ..., a', 0'', ..., a'' \}
\eea
and contains 33 elements. It is straightforward, though tedious, to
enumerate all simplices in this complex. The number of each simplex
type in this simplicial complex for $RP^{3}\times S^{1}$ is,
\bea
& & 0-simplices \;\;\;\;\;\;\;\;  33 \label{p3num} \\
& & 1-simplices \;\;\;\;\;\;\;\;  339 \nonumber \\
& & 2-simplices \;\;\;\;\;\;\;\;  1026 \nonumber \\
& & 3-simplices \;\;\;\;\;\;\;\;  1200 \nonumber \\
& & 4-simplices \;\;\;\;\;\;\;\;  480 \;\; . \nonumber
\eea

   In order to compute the partition function in these theories, we
need to first determine the admissible field configurations. This
means finding the gauge inequivalent solutions to the equations,
\bea
& & [\delta A ]_{012} = [A_{12} - A_{02} + A_{01}] = 0 \\
& & [\delta B ]_{0123} = [B_{123} - B_{023} + B_{013} - B_{012}]
= 0 \;\; ,\nonumber
\eea
where gauge transformations are given by,
\bea
& & A'_{01} = [ A_{01} - \omega_{1} + \omega_{0}] \\
& & B'_{012} =[ B_{012} - \lambda_{12} + \lambda_{02} - \lambda_{01}]
\;\; . \nonumber
\eea
We remind the reader that the brackets in these equations denote that
the quantity inside is to be taken mod-$p$.

The number of gauge inequivalent solutions to these equations is in
correspondence with the first and second cohomology groups of the
complex $K$ with coefficients in $Z_{p}$. Beginning with the well
known homology groups with integer coefficients for $RP^{3}$ and
$S^{1}$ \cite{JM,Rot} ,
\bea
& & H_{0}(RP^{3}) = H_{3}(RP^{3}) = H_{0}(S^{1}) = H_{1}(S^{1}) = Z \;\; ,
\non \\
& & H_{1}(RP^{3}) = Z_{2} \;\; , \;\; H_{2}(RP^{3}) = 0 \;\; ,
\eea
the Eilenberg-Zilber theorem,
\bea
H_{n}( X \times Y ) = \sum_{i+j=n} \; H_{i}(X) \otimes H_{j}(Y) \oplus
\sum_{p+q=n-1} \; Tor(H_{p}(X),H_{q}(Y))     \;\; ,
\eea
computes the homology of the product $K = RP^{3} \times S^{1}$, and
one finds
\bea
H_{0}(K) = H_{3}(K) = H_{4}(K) = Z\;\; ,\;\;
H_{1}(K) = Z \oplus Z_{2} \;\; ,\;\; H_{2}(K) = Z_{2} \;\; .
\eea
The universal coefficient theorem for cohomology,
\bea
H^{n}(X,G) = Hom( H_{n}(X),G) \oplus Ext( H_{n-1}(X),G ) \;\; ,
\eea
then gives the required cohomology groups which for $G=Z_{p}$ are,
\bea
H^{0}(K,Z_{p}) = H^{4}(K,Z_{p}) = & & Z_{p} \\
H^{1}(K,Z_{p}) = H^{3}(K,Z_{p}) = & & \left\{ \begin{array}{ll}
                    Z_{p}              & \mbox{for $p$ odd} \non \\
                    Z_{p} \oplus Z_{2} & \mbox{for $p$ even} \non \\
                    \end{array}
                    \right.  \\
H^{2}(K,Z_{p}) = & &   \left\{  \begin{array}{ll}
                    0                  & \mbox{for $p$ odd} \non \\
                    Z_{2} \oplus Z_{2} & \mbox{for $p$ even.} \non \\
                    \end{array}
                    \right.
\eea

   Now, the partition function in this four dimensional
model essentially amounts to a
sum over field configurations which represent inequivalent classes in
the cohomology groups $H^{1}(K,Z_{p})$ and $H^{2}(K,Z_{p})$.
While the scale factor we have introduced, $1/G^{N_{1}}$,
defines a subdivision invariant quantity, one could also adopt
a different normalization where the partition function is precisely
proportional - in a way independent of the simplicial complex - to a
sum over these classes. To relate our original partition function to the
latter, we need to count carefully gauge equivalent copies of all field
configurations. Also, knowing the number of allowed gauge constraints
is important for the purposes of explicitly finding all solutions.

   For link based fields, the counting of gauge copies is the same as
in lattice gauge theory where a different copy of the gauge group is
assigned to each vertex. The gauge transformation here is,
\bea
A^{\prime} = [ A - \delta \, \omega ] \;\; ,
\eea
and what we seek is the dimension of the image of the map
\bea
\delta^{0} : C^{0}(K) \rightarrow C^{1}(K) \;\; .
\eea
Here we have explicitly attached a superscript to $\delta$ to denote the
restriction to $C^{0}$. But $\delta^{0}(C^{0})$ is isomorphic to,
\bea
C^{0} / Ker( \delta^{0} ) \;\;
\eea
and the kernel of $\delta^{0}$ is 1 dimensional for a connected complex
(same as $H^{0}(K,Z_{p})$). One then sees that the image of this map
has dimension $N_{0} - 1$, and this is the number of links we can gauge
fix.

  For the 2-simplex field $B$, the counting is only slightly more difficult.
As a gauge field, we are assigning an element in the gauge group to each link,
and we seek the dimension of the image of the map,
\bea
\delta^{1} : C^{1}(K) \rightarrow C^{2}(K)\;\; ,
\eea
which will tell us how many of the 2-simplex fields can be gauged away. Let
us restrict the following discussion to the case of $p$ a prime number so
$Z_{p}$ is a field. The kernel of $\delta^{1}$, the 1-cocycles,
is then parametrized
by the image of $\delta^{0}$ together with $H^{1}(K,Z_{p})$. When p is prime,
the later cohomology group is then a sum of copies of $Z_{p}$; let
$h^{1}$ denote the number of these copies, i.e. the dimension of
$H^{1}(K,Z_{p})$ as a vector space over $Z_{p}$. Hence the image of
$\delta^{1}$ has dimension,
\bea
N_{1} - (N_{0} - 1) - h^{1} \;\; . \label{btree}
\eea
Putting these numbers for the maximal trees together, we can, for $|G|=p$
a prime number, write the partition function (\ref{z4dim}) as
\bea
Z = \frac{1}{|G|^{h_{1}} } \;\; \sum_{flat'} \;\; W[K] \;\;,
\eea
where $flat'$ indicates the sum is over the gauge
inequivalent configurations which are the cohomology classes.

  In the case of the complex for $RP^{3}\times S^{1}$, we have
1026 equations for the $A$ field and 1200 for $B$. These are highly
redundant due to the Bianchi identities, but nevertheless, the number
is quite large and one needs to make maximal use of the gauge freedom.
For the link based gauge field $A$, one is allowed to set to zero
(gauge fix) the
fields on a maximal tree. A maximal tree is any maximal set of links
which contains no closed loops, and as we saw above, that number is
always one less than
the number of vertices. It is trivial to pick such a set by inspection,
and in this case we can gauge fix the $A$ field on 32 1-simplices.

For the 2-colour field $B$, the situation is somewhat more intricate.
In practice, it is not easy to
identify such a set by inspection of the complex. Instead, we solved
this problem by associating a vector of length 339 (one place for each
link field $\lambda$) to each of the 1026 2-simplices; this vector
then represents
a gauge transformation. By using a program in Mathematica
\cite{Wolf}, we could
find a maximal number of linearly independent vectors which was
found to be 306 for our complex. There is one additional complication
however. The RowReduce routine in Mathematica gives vectors which are
linearly independent over the real numbers, and we seek a set which is
linearly independent over $Z_{p}$. One indeed finds a single vector
in that set which is not linearly independent for all $p$. Our gauge choice
then amounts to 305 conditions. This is the number given by (\ref{btree})
for our complex  (\ref{p3num}) when $p=2$. While it is not a maximal
tree in general, it is an allowed choice for all $p$.

Another check on the gauge choice we have made here is to nominate one
link variable as the independent one for each of the gauge conditions
we seek to impose. One then shows that the 305 choices found previously
can be made with no duplications.

At this stage, solving the equations subject to a maximal gauge choice
is not difficult, though it is somewhat tedious. Typically, repeated use of
the gauge conditions forced most other fields to vanish. For the $A$
field, we found that the nonzero pieces could be parameterized in terms
of two mod-$p$ variables $a$ and $x$, where $[ 2\; a ] = 0$ and $x$
was unconstrained.
The same is also true for the $B$ field, where we parametrize the solution
in terms of $b$ and $y$, with $[2 \; b ] = 0$ and $y$ unconstrained.

For each of these field configurations, one then computes the
Boltzmann weight which is a product of 480 factors, one for each 4-simplex
in the complex. We find the Boltzmann weight
\bea
\exp [ \frac{2 \pi i k}{p^{2}}\;2\; b \; a ]\;\;.
\eea
One notices immediately that the Boltzmann weight is independent of
the $x$ and $y$ parameters in the general solution. Since the $p$
odd case has no non-zero solutions for $a$ and $b$, there are no
non-trivial phases. For $p$ even, the sum over
$a$ and $b$ yields
\bea
       3 + (-1)^{k} = 2\cdot 2^{\delta_{2}(k)}\;\;,
\eea
and the partition function at $s = \exp[ 2 \pi i k / p^{2} ]$  is
given by,
\bea
Z[RP^{3}\times S^{1}] = \left\{  \begin{array}{ll}
         2 \cdot 2^{\delta_{2}(k)}        & \mbox{for $p$ even} \\
         1                                & \mbox{for $p$ odd.}  \\
         \end{array}
         \right.
\eea
The symbol $\delta_{p}(k)$ denotes the mod-$p$ delta function; its value
is 1 if $k=0$ mod-$p$, and 0 otherwise.
In detail the calculation for $p$ even takes the form,
\be
Z = \frac{1}{p^{339}}\, p^{305} \, p^{32}\, p^{2} \, (3 + (-1)^{k}) \;\; .
\ee
The number 339 comes from the number of 1-simplices in the complex, the
factors with 305 and 32 take into account the gauge equivalent copies of
the solutions to the flatness equations, and the remaining factors
come from summing the solutions over $a$, $x$, $b$, and $y$.

One can compare this result to that obtained for the 4-sphere $S^{4}$.
In this case, a complex is easily given as the boundary of a single
5-simplex; one has the following data,
\bea
& & 0-simplices \;\;\;\;\;\;\;\;  6 \\
& & 1-simplices \;\;\;\;\;\;\;\;  15 \nonumber \\
& & 2-simplices \;\;\;\;\;\;\;\;  20 \nonumber \\
& & 3-simplices \;\;\;\;\;\;\;\;  15 \nonumber \\
& & 4-simplices \;\;\;\;\;\;\;\;  6   \;\; . \nonumber
\eea
The calculation of the partition function is easily seen to take the form,
\bea
Z[S^{4}] = \frac{1}{p^{15}} \, p^{10} \, p^{5} \, 1 = 1 \;\; ,
\eea
though there are no interesting solutions to the flatness equations and
hence no possibility of non-trivial phases in this case. Hence,
the value of the partition function is independent of $k$.

One triangulates the space $S^{3} \times S^{1}$ in the same manner as for
the projective space we have already considered.
There is no
possibility of any phases and a straightforward calculation gives,
\be
Z[S^{3} \times S^{1} ] = 1 \;\; .
\ee

   It is also straightforward to carry out calculations on the spaces
$L(p,q) \times S^{1}$, where $L(p,q)$ is a lens space \cite{JM}, though
the triangulations \cite{Brehm} get progressively larger. In this
series, $RP^{3}$ appears as $L(2,1)$. We have also done the analogous
computation of the partition function for $L(5,1) \times S^{1}$. We
find that the Boltzmann weight depends on two variables $a$ and $b$
which must satisfy $[5 a] = [5 b] = 0$, so there are no non-unit
phases when $p$ is not a multiple of 5. When $p$ is a multiple of 5,
the Boltzmann weight takes the form
\bea
\exp [ \frac{ 2 \pi i k}{p^{2}} \; 5\; b \; a ] \;\; .
\eea
The partition function becomes,
\bea
Z[L(5,1) \times S^{1}] =   \left\{  \begin{array}{ll}
      5 \cdot 5^{\delta_{5}(k)}       & \mbox{for $p$ a multiple of 5} \\
      1                               & \mbox{otherwise.}  \\
      \end{array}
      \right.
\eea
{}From these examples, we see that there is generally a dependence of the
partition function on the coupling parameter. Since the partition
function on $M_{3} \times S^{1}$ gives the dimension of the Hilbert
space \cite{At,T},
we see that the quantum Hilbert space associated to $M_{3}$
differs from the classical one ($k=0$).

\section{The Bockstein Operator}

   At first sight, the construction of this general class of models
that we are considering in this paper may seem mathematically unorthodox.
The motivation stemmed from a desire to realize discrete Chern-Simons
and BF theories from a concrete point of view. The intuition was that
one should make use of the coboundary operator on simplicial cochains
in some fashion, but we did so with a cup product which differed from
the usual one in so far as we did not take the product mod-$p$. However,
we can, in retrospect, make an observation which brings the whole
construction into orthodoxy, and this is the connection with the
Bockstein operator \cite{DFN}; the homotopy-type nature of
these models is then transparent. We shall restrict attention here
to the case of closed manifolds, so that the fundamental class exists.

   Let $x$ be an element of the simplicial cohomology group
$H^{q}(K,Z_{p})$, and let $\bar{x} \in C^{q}(K,Z)$ denote a representative
of $x$ as an integral cochain. Since $[\delta \, x] = 0$, this means
$\delta \, \bar{x} = p \, u $, for some integral $(q+1)$-cochain $u$.
The Bockstein operator
\be
\beta: H^{q}(K,Z_{p}) \rightarrow H^{q+1}(K,Z_{p})
\ee
is defined by
\be
\beta(x) = [ \frac{1}{p} \; \delta \, \bar{x} ] = [ u ] \;\; .
\ee
In terms of the Bockstein operator and normal cup product, we can then
rewrite the Boltzmann weight (\ref{sdibw}) quite simply as,
\be
\exp [ \frac{2 \pi i k}{p} \; < \, B \cup \beta(A),\; \sigma_{4} \, > ]\;\; ,
\ee
where $\sigma_{4}$ is a 4-simplex.
Of course, the Boltzmann weight for the abelian Dijkgraaf-Witten
\cite{DW} theory can also be so written;
\be
\exp [ \frac{2 \pi i k}{p} \; < \, A \cup \beta(A),\; \sigma_{3} \, > ]\;\; ,
\label{bockdw}
\ee
where the connection with Chern-Simons theory is striking. The key
observation is that the Bockstein operator and not the simple coboundary
operator is what is relevant in the construction of these models with
gauge group $Z_{p}$.

The extension of this theory to 5 and higher dimensions
is then transparent, and one takes the Boltzmann weight,
\be
\exp [ \frac{2 \pi i k}{p} \; < \, A \cup \beta(A) \cup \cdots
\cup \beta(A),\; \sigma_{2 m + 1} \, > ]\;\; ,
\ee
where we have $m$ factors of $\beta(A)$. It is worth noting that theories
whose partition functions lead to Gauss sums such as the 3d Dijkgraaf-Witten
theory will also appear in 7d where the link field $A$
in (\ref{bockdw}) becomes a 3-colour field. In this case, partition functions
will generally be complex valued unlike the the class of $B \cup \beta(A)$
theories where (for $p$ prime at least) they are real.

\section{Manifolds with Boundary}

    A general axiomatic framework was presented for
topological quantum field theory ($TQFT$) in \cite{At};
see also \cite{T,Y2,Wak}. As we have seen, the
general class of abelian models considered here can be expressed in terms
of standard modulo-$p$ cohomological operations, and one expects
the axioms of $TQFT$ to be satisfied by these models.

  Let $K$ be a 4-manifold with boundary $\partial K$. The partition
function of these models is well defined, and represents a transition
amplitude when we specify the field configurations on the boundary
components. We take the field configurations on the boundary to be
flat, and define the allowed field configuration on $K$ to be all
flat configurations which extend those specified on the boundary. The
partition function remains subdivision invariant as long as we keep
the triangulation on the boundary fixed \cite{Wak}. It is convenient,
however, to rescale the partition function by
\bea
Z'[K] = w^{N_{1}(\partial K) } \; Z[K]\;\; ,
\eea
where $w=\sqrt{|G|}$, and $N_{1}(\partial K)$ is the number of
1-simplices on the boundary
$\partial K$. This scaling gives the following gluing rule,
\be
Z'[K,\tau_{1},\tau_{2}] = \sum_{\tau_{3}} \;\;
Z'[K_{1},\tau_{1},\tau_{3}] \cdot Z'[K_{2},\tau_{3},\tau_{2}] \;\; .
\ee
Here $K_{1}$ is a cobordism between boundary manifolds $\Sigma_{1}$
and $\Sigma_{3}$ with fixed flat field configurations $\tau_{1}$
and $\tau_{3}$, and similarly for $K_{2}$. $K$ represents a composition
of $K_{1}$ and $K_{2}$ and the above sum is over all intermediate
flat field configurations.

Consider now a gauge transformation
of the $B$ field, when a boundary $\partial K$ is present.
{}From (\ref{bwgt}), we see that the Boltzmann weights are
related by a phase factor depending only on the boundary values of
the fields, namely:
\be
s^{<B^{\prime} \cup \delta A,\, K>}= s^{<B \cup \delta A,\,K>}
s^{- <\lambda \cup \delta A,\,\partial K>} \;\;.
\ee
When computing the partition function on $K$, we sum over all
allowed field configurations with fixed boundary data, and thus we
see that the partition function also transforms with this phase
factor. It is equally simple to determine the behaviour under
a gauge transformation of the $A$ field.

For the purposes of illustration let us consider the case of a $4$-manifold
of the form $M_3 \times I$, where $I$ is the unit interval, and $M_{3}$
is some bounding $3$-manifold. The value of the partition function
then gives a transition amplitude between the two copies of
$M_{3}$. Given that $H^{*}(M_{3}\times I,Z_{p}) = H^{*}(M_{3},Z_{p})$, we
know that the transition matrix $Z'_{if}[M_{3}\times I]$ must be diagonal.
Moreover, because of subdivision invariance,
\be
Z'_{if}[M_{3}\times I] = \sum_{j}\; Z'_{ij}[M_{3}\times I] \cdot
Z'_{jf}[M_{3}\times I]\;\; ,
\ee
which shows that any diagonal element can only be 0 or 1.
Since the two copies of the bounding manifold appear
with opposite orientation, the behaviour of the partition
function under changes of cohomology representatives (gauge transformations)
is given by
\be
Z'_{i^{\prime}\,f^{\prime}} = \exp[i \alpha_{i}] \,Z'_{i\,f} \,
\exp[- i \alpha_{f}] \;\;,
\ee
where $Z'_{i\,f}$ denotes the transition amplitude between the initial
and final copy of $M_{3}$.
It is then clear, for example, that the diagonal elements along with the trace
of $Z'_{i\,f}$ are gauge invariant quantities.
In particular, we have the result
\be
\sum_{i}\; Z'_{i\,i}[M_{3}\times I] = Z[M_{3} \times
S^{1}]\;\;,  \label{trace}
\ee
which we can interpret as the dimension of the Hilbert space associated
to $M_{3}$. In taking the above trace, one must
take into account the gauge equivalent copies
of the fixed boundary data.

Indeed, the value of the partition function for the manifold
$M_{3} \times S^{1}$ can be obtained more easily by first
performing the computation on $M_{3} \times I$, and then taking the trace.
In this way, one sees that a vertical tower (see (\ref{tower})) of
only four $4$-simplices is required.

  Up to gauge equivalence, field configurations on any single boundary
component $M_{3}$ are in one to one correspondence with the set
$H^{1}(M_{3}, Z_{p}) \times H^{2}(M_{3} , Z_{p})$.
We can define a complex vector space $V(M_{3})$ associated to $M_{3}$
by taking it to be the vector space freely generated by this set of
field configurations. This we might call the classical Hilbert
space of $M_{3}$ \cite{DW}. However, the partition function on the
cylinder represents a transition amplitude, and the map
$Z'_{if}[M_{3}\times I]$
may well have a non-zero kernel. The quantum Hilbert space $H(M_{3})$
is defined to be,
\be
H(M_{3}) = V(M_{3}) / Ker(Z'_{if}[M_{3}\times I]) \;\; ,
\ee
and its dimension is given by the above trace (\ref{trace}). Thus,
our computations show that the quantum Hilbert space of these
models is generically different from the classical Hilbert space.
By classical Hilbert space one simply refers to the situation where
$k=0$ and the kernel of $Z'_{if}$ on the cylinder always vanishes.

  It is interesting to go further and identify precisely the zero modes
in the examples we have computed. For the $p=2$, $k=1$ theory with $RP^{3}$
boundaries, the calculation of the partition function on $RP^{3}\times
S^{1}$ indicates a Hilbert space of dimension 2, compared to the classical
$k=0$ result of 4. Hence there must be two zero modes in the ``propagator''
on the cylinder. One finds that these zero modes correspond to the
non-trivial $H^{1}(RP^{3},Z_{2})$ configuration which means that the Hilbert
space is in correspondence with $H^{2}(RP^{3},Z_{2})$. Whether this is
a general phenomenon, or is something peculiar to the lens spaces, is
not known.

Let us now examine the behaviour of the partition function with respect
to the connected sum of manifolds. For manifolds $M_{1}$ and $M_{2}$,
the connected sum is denoted by $M = M_{1} \# M_{2}$.
The manifold $M$ is produced by first excising a $4$-ball from
each of the components $M_{1}$ and $M_{2}$, which are then identified
along their common $S^{3}$ boundary.
According to the tenets of the axiomatic
approach, the partition function on a closed manifold
can be computed by cutting the manifold along a common
boundary, and then taking the pairing between the state
vectors in the dual Hilbert spaces.

Now, in order to obtain a relationship
between the partition function $Z[M]$ and that of its components
$Z[M_{1}]$ and $Z[M_{2}]$, we recall the value of the partition function
$Z[S^{3} \times S^{1}] = 1$. We thus see that the $S^{3}$ Hilbert space is
$1$-dimensional.
It is then a simple consequence of $1$-dimensional linear algebra \cite{W}
to see that the following relationship holds:
\be
Z[M_{1} \# M_{2}]\,Z[S^{4}]  = Z[M_{1}]\, Z[M_{2}]\;\;.
\ee
Since $Z[S^{4}] = 1$ for the four dimensional model under consideration,
the partition function behaves multiplicatively under connected sum.

\section{Concluding Remarks}

The four dimensional model we have considered here is part of a generic
construction available in all dimensions \cite{BR}. The cornerstone of
these state sum models is a partition function which is a sum over
simplicial cohomology classes of a certain Boltzmann weight. This phase
factor arises from the modulo-$p$ valued intersection form between
those classes. In particular, we have seen that the relevant ``kinetic"
operator used to define these actions is provided by the Bockstein
coboundary operator. We remark that one can also consider the ``non-kinetic"
type models \cite{BR2} as providing observables for the theories
presented here.
In three dimensions, it reduces to the Dijkgraaf-Witten model \cite{DW}
which is related to group cohomology. Such a connection is not transparent
in general.

We have shown here that the four dimensional model is indeed
non-trivial in the sense that interesting phases can be obtained; without
them the model only counts cohomology classes.
In this regard, one sees that the dimensions of the quantum Hilbert
spaces are in general different from the classical dimensions.
By construction, these models lead to piecewise linear invariants;
however, with the insight that they can be formulated (in the closed
case) in terms of the Bockstein operator, one sees that the models
yield homotopy-type invariants.

In \cite{CY}, a four dimensional subdivision invariant model was described
in terms of combinatorial data.
Subsequently, it was established in \cite{R,CKY}, that the partition function
was expressible in terms of the Euler and Pontryagin numbers, and as
such encoded classical topological data.
On the other hand, this model could then be viewed  as describing
classical invariants in terms of a quantum state sum.
The models presented here can similarly be viewed as
providing a quantum state sum formulation of
classical modulo-$p$ cohomological data.
Perhaps it is also worth remarking on how these models differ
from the structures presented in \cite{AS}.
Quite apart from having to address issues of
regularizing the formally divergent path integrals of those models,
one is also dealing with cohomology with real coefficients;
as such, the models are insensitive to the presence of
any torsion subgroups.
However, as we have seen for the models discussed here, the essence
of non-triviality lies in the presence of torsion in the cohomology
groups.

{\bf Acknowledgements}\\
M.R. would like to thank Ron Kantowski for computer access
at the University of Oklahoma.


\begin{thebibliography}{99}
\bibitem{BR} D. Birmingham and M. Rakowski, {\em On Dijkgraaf-Witten
Type Invariants}, University of Amsterdam preprint, ITFA-94-07, February
1994, hep-th/9402138.
\bibitem{DW} R. Dijkgraaf and E. Witten, {\em Topological Gauge Theories
and Group Cohomology}, Commun. Math. Phys. 129 (1990) 393.
\bibitem{JM} J. Munkres, {\em Elements of Algebraic Topology},
Addison-Wesley, Menlo Park, 1984.
\bibitem{Rot} J. Rotman,  {\em An Introduction to Algebraic Topology},
Springer-Verlag, New York, 1988.
\bibitem{Still} J. Stillwell, {\em Classical Topology and Combinatorial
Group Theory}, Springer-Verlag, New York, 1980.
\bibitem{Y1} D.N. Yetter,  {\em State-sum Invariants of
3-Manifolds Associated to Artinian Semisimple Tortile Categories},
Top. and its App. 58 (1) (1994) 47.
\bibitem{CY} L. Crane and D.N. Yetter,  {\em Categorical
Construction of 4D Topological
Quantum Field Theories}, in {\em Quantum Topology}, L.H. Kauffman and
R.A. Baadhio, eds., World Scientific (1993) 120.
\bibitem{Alex} J.W. Alexander, {\em The Combinatorial Theory of
Complexes}, Ann. Math. 31 (1930) 292.
\bibitem{Pachner} U. Pachner, {\em P.L. Homeomorphic Manifolds are
Equivalent by Elementary Shelling}, Eur. J. Comb. 12 (1991) 129.
\bibitem{Felder} G. Felder and O. Grandjean, {\em On Combinatorial
Three-Manifold Invariants}, in {\em Low-Dimensional Topology and
Quantum Field Theory}, ed. H. Osborn, Plenum Press, New York, 1993.
\bibitem{Brehm} U. Brehm and J. Swiatkowski, {\em Triangulations
of Lens Spaces with Few Simplices}, T.U. Berlin preprint, 1993.
\bibitem{Wolf} S. Wolfram, {\em Mathematica}, Addison-Wesley, New York,
1988.
\bibitem{At} M.F. Atiyah, {\em Topological Quantum Field Theories},
Publ. Math. IHES 68 (1989) 175.
\bibitem{T} V.G. Turaev, {\em Quantum Invariants of 3-Manifolds},
{\em Publ. de l'Institue de Recherche Math\'{e}matique Avanc\'{e}e}
509/P-295 CNRS, Strasbourg, France (1992).
\bibitem{DFN} B. Dubrovin, A. Fomenko, and S. Novikov, {\em Modern Geometry
- Methods and Applications, Part III}, Springer-Verlag, New York, 1990.
\bibitem{Y2} D.N. Yetter, {\em Triangulations and TQFT's}, in
{\em Quantum Topology}, L.H. Kauffman and R.A. Baadhio, eds.,
World Scientific, (1993) 354.
\bibitem{Wak} M. Wakui, {\em On Dijkgraaf-Witten Invariant for 3-Manifolds},
Osaka J. Math. 29 (1992) 675.
\bibitem{W} E. Witten, {\em Quantum Field theory and the Jones
Polynomial}, Commun. Math. Phys. 121 (1989) 351.
\bibitem{BR2} D. Birmingham and M. Rakowski, {\em Discrete Quantum
Field Theories and the Intersection Form}, Mod. Phys. Lett. A9 (1994) 2265.
\bibitem{R} J.D. Roberts, {\em Skein Theory and Turaev-Viro
Invariants}, Cambridge University preprint, 1993.
\bibitem{CKY} L. Crane, L.H. Kauffman, and D. Yetter, {\em Evaluating
the Crane-Yetter Invariant}, in {\em Quantum Topology}, L.H. Kauffman
and R.A. Baadhio, eds., World Scientific, (1993) 131.
\bibitem{AS} A.S. Schwarz, {\em The Partition Function of a Degenerate
Quadratic Functional and the Ray-Singer Invariants}, Lett. Math.
Phys. 2 (1978) 247.
\end{thebibliography}
\end{document}